\documentclass[reprint,amsmath,amssymb,aps,PRL]{revtex4-2}

\usepackage{graphicx}
\usepackage{dcolumn}
\usepackage{color,soul}
\usepackage{bm}
\usepackage{natbib}
\usepackage{ulem}
\usepackage{pgfplots}
\pgfplotsset{compat=1.18}
\usepackage{amsmath}
\usepackage[caption=false]{subfig}
\usepackage{multirow}

\begin{document}

\preprint{APS/123-QED}



\author{G. Scamps}
\author{A. Guilleux}
\affiliation{Université de Toulouse, CNRS/IN2P3, L2IT, Toulouse, France}

\author{D. Regnier}
\author{A. Bernard}
\affiliation{CEA, DAM, DIF, 91297 Arpajon, France}
\affiliation{Université Paris-Saclay, CEA, LMCE, 91680 Bruyères-le-Châtel, France}


\date{\today}

\title{Uncertainty Principle and Angular Momentum Generation in Microscopic Fission Models}


\begin{abstract}
The generation of angular momentum (intrinsic spin) in fission fragments has recently attracted renewed attention. While several microscopic approaches reproduce the spin distribution qualitatively using projection techniques, the physical origin of the fragments’ angular momentum in density functional theory remains unclear. In this work, we investigate the mechanisms responsible for the spin distribution of fission fragments within a microscopic TDDFT framework. We compare spin distributions obtained from projection operators with those predicted by a simple expression derived from the uncertainty relation between angle and angular momentum, where angular fluctuations are estimated using a Monte Carlo sampling of nucleon positions. We find that a large portion of the spin distribution obtained from projection methods can be explained by the uncertainty principle. Our results thus show that, within microscopic approaches, the spin of fission fragments originates primarily from quantum uncertainty associated with their orientation angle with respect to the fission axis, mainly due to quadrupole deformation and, to a lesser extent, octupole deformation.
\end{abstract}

\maketitle


The uncertainty principle, first formulated by Heisenberg~\cite{heisenberg1927anschaulichen} and later generalized by Robertson~\cite{robertson1929uncertainty}, expresses a fundamental quantum limitation on the simultaneous knowledge of conjugate variables. Beyond the well-known position–momentum relation, this principle also applies to angular observables, linking the dispersion of angular momentum to that of the orientation angle. The uncertainty principle not only limits measurement precision but also fundamentally characterizes the nature of quantum states. In particular, the angle–angular momentum uncertainty relation~\cite{Franke-Arnold_2004} implies that a well-defined orientation of a system necessarily leads to a finite spread in its angular momentum components, typically following a spin cut-off distribution. A rare case of phenomena where that uncertainty principle play a role between spin and angle is related to the polarisation of photons~\cite{Franke-Arnold_2004}.
 In the case of fission fragments, their intrinsic deformation defines an orientation with respect to the fission axis; consequently, quantum uncertainty in this orientation naturally gives rise to a distribution of angular momenta, providing a macroscopic origin for the fragments’ intrinsic spin~\cite{MIKHAILOV:1999,Bonneau2007,Scamps:2023}.

The characterization of fission fragment properties is a major goal in modern studies of nuclear fission \cite{Bender_2020, SCHUNCK2022103963, Schunck:2016}. Among these properties, the angular momentum (or spin) of the fragments has attracted particular interest following recent experimental results \cite{Wilson:2021,travar2021experimental}, which indicate that fragment spins exhibit a saw-tooth pattern. This pattern is often interpreted as reflecting a dependence on the deformation of the fragments. These new findings have renewed interest in both experimental \cite{Marin:2021,Giha:2023,Francheteau2024,Marin2024,Cannarozzo:2025} and theoretical studies.

Theoretical macroscopic approaches to the generation of angular momentum in fission fragments can be broadly divided into two classes. The first consists of statistical approaches \cite{Moretto:1980,Dossing:1985,Randrup:2021,Randrup:2022a,Vogt:2021}, in which fluctuations accumulate during the shape evolution from the initial configuration to scission, leading to rotational excitation of the fragments. The second consists of quantum approaches \cite{RASMUSSEN1969465,MIKHAILOV:1999,Bonneau2007,Scamps:2022,Scamps:2023,Shneidman:2025,Shneidman:2002,zhou:2023}, where quantum fluctuation, sometime related to the uncertainty principle is invoked as the source of fragment spin. The first  formulation of this idea was given by Rasmussen, Nörenberg, and Mang \cite{RASMUSSEN1969465}, who established the connection between the angular localization of the fragment wave function and the resulting angular-momentum distribution.

Another class of methods relies on projection techniques applied to static \cite{Bertsch:2019,Marevic:2021,marevic:2025} or time-dependent \cite{Bulgac:2021,Scamps:2023a,Scamps2024} density-functional theory (DFT) descriptions of fission. While these microscopic approaches provide powerful tools for modeling fission, their interpretation is often less straightforward. In particular, in recent works using projection operators, the generation of angular momentum is usually discussed only in terms of a qualitative connection with fragment deformation, without addressing its fundamental origin.

In this Letter, we focus on understanding the microscopic origin of angular momentum generation within DFT-based descriptions of fission, and restrict our discussion to the intrinsic fragment spin at scission, prior to neutron and $\gamma$ emission.

To this end, we use the Gogny-TDHFB code~\cite{hashimoto:2012,Hashimoto:2013,hashimoto2016gauge,scamps2017b,Scamps:2019,regnier2018microscopic} that solve the time-dependent  Hartree-Fock-Bogoliubov equation with the Gogny D1S interaction in a hybrid basis consisting of two-dimensional harmonic oscillator eigenfunctions and one-dimensional spatial mesh (the z-axis or fission axis). A mesh parameter $\Delta z=0.8$ fm is used with $N_z=52$ points. The oscillator wave function are restricted to $n_x+n_y\leq N_{shell}$ with $N_{shell}=9$ and a harmonic oscillator parameter $\hbar \omega=8$~MeV. The time propagation is solved with the Runge-Kutta method in the fourth order with a time-step $\Delta t=2.10^{-3}$ zs. While a quantitative description of fission within TD-DFT generally requires averaging over several dynamical trajectories, the purpose of the present discussion is to provide physical insight rather than statistical observables. In this context, a single representative trajectory is sufficient.

We study three fissioning actinides, $^{230}$Th, $^{240}$Pu, and $^{250}$Cf, which span a region of the nuclear chart where fission is asymmetric and the heavy fragment is influenced by octupole deformation~\cite{Scamps:2018}. These three systems also cover the range in which the characteristic saw-tooth pattern of the fragment spin has been observed experimentally~\cite{Wilson:2021}.

 \textit{Uncertainty principle}—In contrast to the well-known uncertainty relation between position and momentum, several conceptual difficulties arise in formulating an uncertainty principle between the orientation of the system and its angular momentum~\cite{Franke-Arnold_2004}. The first complication is that angular coordinates are periodic variables, which makes it nontrivial to define both their average values and standard deviations in a general sense~\cite{Franke-Arnold_2004,CARRUTHERS1968,Judge1963,Barnett1990}. To overcome this issue, we restrict our analysis to well-oriented systems exhibiting axial symmetry around the $z$-axis and small fluctuations in the orientation of their principal axes. This situation occurs, for example, in nuclear fission, where the fragments are well aligned along the fission axis.

The corresponding uncertainty relation can then be written as
\begin{align}
\Delta \theta\, \Delta L_x > \frac{1}{2},
\end{align}
where $L_x$ denotes the projection of the angular momentum along the $x$-direction expressed in units of $\hbar$ and $\theta$ is the angle between the principal axis of the system and the z-axis. An analogous relation holds for $L_y$.

For a Gaussian wave packet in terms of $\theta$ and the azimutal angle $\varphi$ of the principal axis,
\begin{align}
\Psi(\theta, \varphi) = {\cal N}\,
\exp\!\Big[-\frac{\theta^2}{4\sigma_\theta^2}\Big],  \label{Gausian_wave_packet}
\end{align}
the corresponding spin–cutoff distribution takes the form
\begin{align}
P(L) = \frac{2L+1}{\cal Z}\,
\exp\!\Big[-\frac{L(L+1)}{2\sigma_L^{2}}\Big],  \label{eq:spin_cut_off}
\end{align}
where ${\cal N}$ and ${\cal Z}$ are normalization factors, and the widths satisfy the Heisenberg-type relation (for an intelligent state)
\begin{align}
\sigma_{\theta}\, \sigma_L = \frac{1}{2}. \label{eq:uncert_principle}
\end{align}
This simple relation illustrates that an object oriented with an angular uncertainty of about $1^\circ$ corresponds to a spin cut-off distribution with an average angular momentum of approximately $29\,\hbar$.

\begin{figure}[!t]
\includegraphics[width=\linewidth]{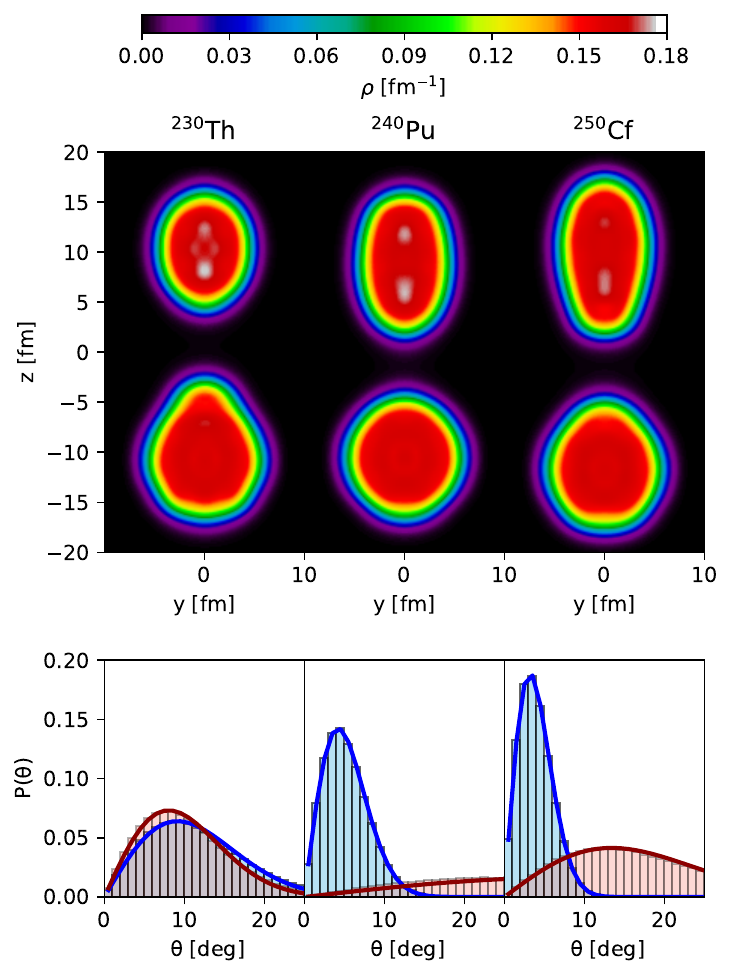}
\caption{\label{fig:density_angular_distribution} 
Top: Slice of the density at $x=0$ at the time when the surface of the fragments are separated by a distance of 6~fm for each fissioning system. 
Bottom: Corresponding angular distribution obtained by the procedure described in the text for the light fragment (blue) and heavy fragment (red). 
The fitted curve from eq.~\ref{eq:gauss_sin} is also shown for each fragment. }
\end{figure}

The second difficulty is the composite nature of nuclei. 
In collective Hamiltonian model such as \cite{Bonneau2007,Scamps:2023,Shneidman:2025} the angle wave packet is  well defined. However, in microscopic model, the definition of the fluctuation of the orientation angle is much more challenging. As seen on Fig.~\ref{fig:density_angular_distribution}, the deformed fragments are oriented along the fission axis. In general, the main orientation can be obtained by computing the principle axes of the one-body density, but is not related to an operator and so  there is no trivial way of obtaining the fluctuations of that orientation.

Here we propose to estimate the distribution of the principal axis orientation going beyond the one-body density picture.
To do so, we sample the position and intrinsic spin of the neutrons and protons from the many-body density of the considered 
Bogoliubov vacuum projected on their good particle number.
These calculations were performed with the code {\tt NucleoScope} implementing a Markov Chain Monte Carlo algorithm to generate
a representative sample of this large dimensional probability distribution \cite{Nucleoscope}.
For each event of this sample, we determine the principle axis of deformation of each fragment and deduce their $\theta$ angle to the z-axis (See more detail in the supplementary material).
This method produces the whole distribution of probability for the $\theta$ angle from which we extract not only the expectation value $\theta$ (which is trivially 0 due to the symmetries) but also its standard deviation $\sigma_{\theta}$. As discussed later, this geometric definition of the fragment orientation accounts only for the orientation associated with quadrupole deformation and neglects additional orientation effects induced by higher-order deformations.

The probability distribution of the angle $\theta$ is shown in the bottom panel of Fig.~\ref{fig:density_angular_distribution} for each nucleus, corresponding to the configurations displayed in the top panel. Each distribution is fitted with the function
\begin{align}
f(\theta) = {\cal N}^2 \sin(\theta)\,
\exp\!\Big[-\frac{\theta^2}{2\sigma_\theta^2}\Big], \label{eq:gauss_sin}
\end{align}
which corresponds to a Gaussian wave packet of the form given in Eq.~\eqref{Gausian_wave_packet}. The good agreement between the calculated distributions and the fitted function indicates that the angular fluctuations are Gaussian. Moreover, the larger the quadrupole deformation of the fragment, the narrower the corresponding angular distribution becomes.

\begin{figure}[!t]
\includegraphics[width=\linewidth]{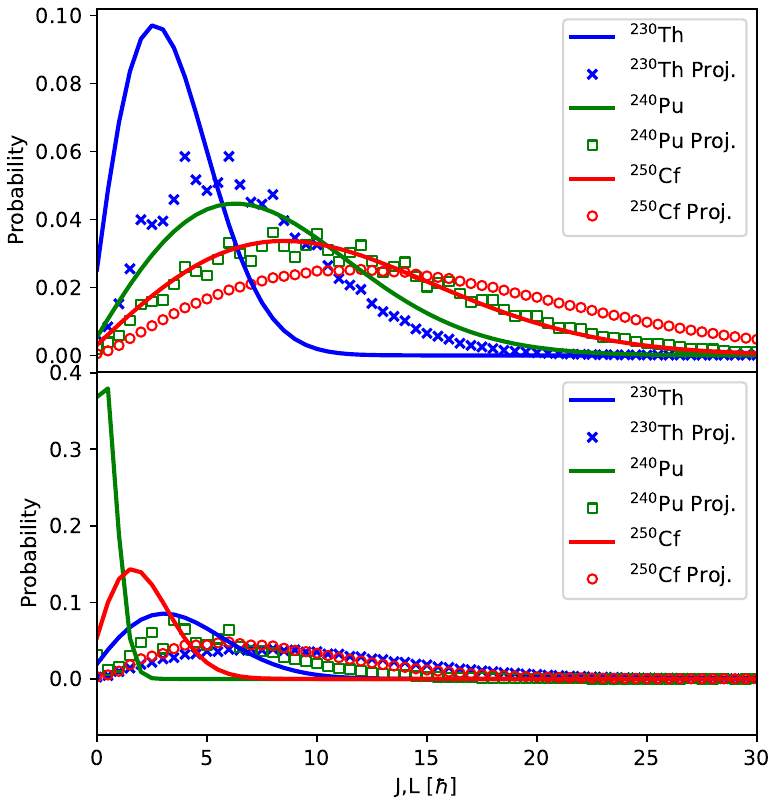}
\caption{\label{fig:spin_l_distributions} Spin distribution obtained from the projection (eq.~\ref{eq:Prob_JK}) compared to the spin cut-off formula (eq.~\ref{eq:spin_cut_off}) assuming the uncertainty principle (eq.~\ref{eq:uncert_principle}) for the light (top) and heavy (bottom) fragments. }
\end{figure}

From the fluctuations of the $\theta$ angle and the uncertainty relation of Eq.~\eqref{eq:uncert_principle}, we can estimate spin cut-off parameters $\sigma_L$, whose values are listed in Table~\ref{tab:sigma}. 
The corresponding angular momentum distributions based on Eq.~\ref{eq:spin_cut_off} are shown in Fig.~\ref{fig:spin_l_distributions}.

\begin{table}[h!]
\centering
\begin{tabular}{|c|c|c|c|c|}
\hline
Nucleus & Fragment & $\sigma_L$ (unc. princ.) & $\sigma_J$ (overlap) & $\sigma_J$ (proj) \\ \hline
\multirow{2}{*}{$^{230}$Th} & L & 2.90 & 5.74 & 5.81 \\ 
                             & H & 3.37 & 7.84 & 7.88 \\ \hline
\multirow{2}{*}{$^{240}$Pu} & L & 6.79 & 9.32 & 9.37 \\ 
                             & H & 0.75 & 4.67 & 4.93 \\ \hline
\multirow{2}{*}{$^{250}$Cf} & L & 9.00 & 12.18 & 12.27 \\ 
                             & H & 2.12 & 6.50 & 6.63 \\ \hline          
\end{tabular}
\caption{\label{tab:sigma}   Spin cut-off parameter for each fissioning nucleus and fragment deduced from the angle fluctuation coupled with the uncertainty principle ($\sigma_L$), from the overlap values ($\sigma_J$), and from the exact projections ($\sigma_J$).}
\end{table}

\begin{figure}[!t]
\includegraphics[width=\linewidth]{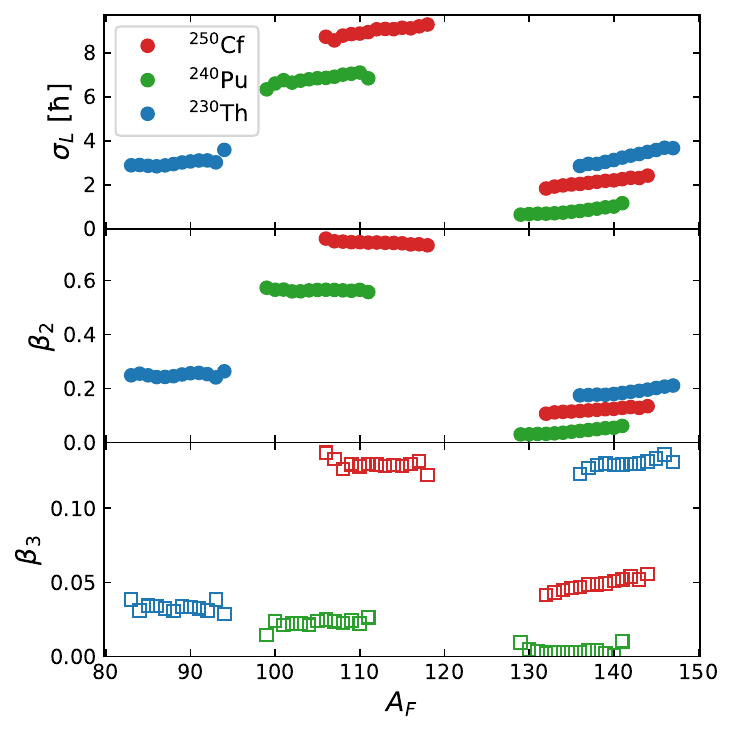}
\caption{\label{sigmaJ_beta_AF}  Fragment-mass dependence of the angular-momentum width $\sigma_L$ (top) and the quadrupole and octupole deformation parameters $\beta_2$ and $|\beta_3|$ (bottom) for the fissioning nuclei $^{230}$Th, $^{240}$Pu, and $^{250}$Cf. Only fragments with more than 300 events are included. The angular-momentum widths $\sigma_J$ are extracted from the folded angular distributions of the fragments, whereas the deformation parameters are computed from the quadrupole (solid symbols) and octupole (open symbols) moments of the fragment nucleon distributions.  
}
\end{figure}

When sampling the positions of the nucleons, it is possible to keep only events giving a specific mass split between the fragments.
Computing the probability distribution of $\theta$ on such a subset of events yields the standard deviation $\sigma_L$ 
as a function of the mass split that we plot in the top panel of Fig.~\ref{sigmaJ_beta_AF}.
The middle and bottom panels show respectively the quadrupole and octupole deformation parameters, computed as
\begin{align}
\beta_{\lambda} = \frac{4\pi}{3A\,(r_0 A^{1/3})^{\lambda}}\, Q_{\lambda 0},
\end{align}
with $r_0 = 1.2$~fm, and
\begin{align}
Q_{20} &= \sqrt{\frac{5}{16\pi}} \int d^3r\, \rho(\mathbf{r}) (2z^2 - x^2 - y^2),\\
Q_{30} &= \sqrt{\frac{7}{16\pi}} \int d^3r\, \rho(\mathbf{r}) [2z^3 - 3z(x^2 + y^2)],
\end{align}
where $\rho(\mathbf{r})$ is the nucleon density.
This figure shows that the saw-tooth pattern of the spin distribution can be understood from the uncertainty principle: a clear correlation emerges between the quadrupole deformation parameter $\beta_2$ and the spin cut-off parameter $\sigma_L$.

\textit{Projection}— The exact way of deducing the spin distribution from a HFB state $| \Psi \rangle$ is to use the projection operator \cite{Scamps:2023a,ZHANG2025139828},
\begin{align}
&P(J_F) = \sum_{K_F} \langle \Psi | \hat P_{K_F K_F}^{J_F} | \Psi \rangle, \label{eq:Prob_JK}
\end{align}
with the operator,
\begin{align}
&\hat P^{J}_{MK} = \frac{(2J+1)}{16\pi^2} \int \!\!d\Omega {\cal D}^{J*}_{MK} (\Omega) \,
e^{i\alpha \hat J_z} e^{i\beta \hat J_y} e^{i\gamma \hat J_z}, \label{eq:PJMK}
\end{align}
with $\Omega=(\alpha,\beta,\gamma)$ representing a separate set of the three Euler angles  corresponding to a rotation of each fission fragment and $ {\cal D}^{J*}_{MK} (\Omega)$ the Wigner D-matrix. The overlaps are determined using the Pfaffian method described in Ref.~\cite{Bertsch:2012,Scamps:2013}.  
To perform the rotation, the wave functions are first expanded from the hybrid basis onto a three dimensional grid with dimensions $n_x = n_y = 30$ and $n_z = 52$. The rotations are then performed exactly using the Lagrange mesh method (see more detail in the supplementary material).
The resulting distribution is shown in Fig.~\ref{fig:spin_l_distributions} and the corresponding spin cut off parameter is listed in Table~\ref{tab:sigma}.
We see in this table that the uncertainty principle, with our method to determine the angular fluctuation, represents the dominant mechanism of angular momentum generation in well deformed fragments.
We estimated, using a Monte Carlo sampler, that the fluctuation of the sum of the intrinsic spins of the nucleons is around $\sigma_s \approx 1 \hbar$, which is insufficient to explain the difference between $\sigma_L$ determined via the uncertainty principle and $\sigma_J$ determined via the projection. We also did the same calculation with the wave-function used in Ref. \cite{Scamps:2023a} with similar results, showing that the present conclusion is not due to the use of the hybrid basis or the Gogny interaction.

 An Heisenberg-type relation also naturally emerges within the projection formalism when the overlap between rotated many-body states can be approximated by a Gaussian,
\begin{align}
 \langle \Psi |  e^{i\alpha \hat J_z} e^{i\beta \hat J_y} e^{i\gamma \hat J_z}  | \Psi \rangle 
 \simeq   \exp\Big[- \frac{\beta^2}{8 \sigma_\beta^2} \Big],  
 \label{Gausian_overlap}
\end{align}
as obtained, for instance, from a Gaussian wave packet in the collective angle, Eq.~\eqref{Gausian_wave_packet}.
This Gaussian behavior reflects the finite angular localization of the many-body wave function in the rotational degree of freedom, characterized by a width $\sigma_\beta$, and does not rely on any geometric definition of the fragment orientation (see also the supplementary material)).
Because the orientation angle and the angular momentum are conjugate variables, this angular localization directly implies a minimal fluctuation of angular momentum.
In this limit, one recovers the standard spin cut-off distribution of Eq.~\eqref{eq:spin_cut_off}, with a width $\sigma_J$ satisfying the uncertainty relation $\sigma_\beta \sigma_J = 1/2$,
which can be shown analytically for small angular fluctuations.

In practice, the full overlap involving three Euler rotations generally deviates from a pure Gaussian form due to the presence of multiple $K$ components, leading
to a more complex dependence on the angles $\alpha$ and $\gamma$ (see Ref.~\cite{Scamps:2023a}). Nevertheless, when the dominant contributions satisfy $|K| \ll J$, the dependence on the additional Euler angles can be effectively reduced, and an equivalent one-angle description provides an accurate approximation of the spin distribution. Such a reduction has been employed in previous microscopic calculations, including to fission dynamics~\cite{Bulgac:2021}.
To assess the validity of the uncertainty-based interpretation in this case, we fit the overlap
$\langle \Psi | e^{i\beta \hat J_y} | \Psi \rangle$
with the Gaussian form of Eq.~\eqref{Gausian_overlap} and compare the extracted $\sigma_J$ with the value obtained from the exact angular-momentum projection (see Table~\ref{tab:sigma}).
To assess the validity of the uncertainty-based interpretation in this case, we fit the overlap
$\langle \Psi | e^{i\beta \hat J_y} | \Psi \rangle$
with the Gaussian form of Eq.~\eqref{Gausian_overlap} and compare the extracted $\sigma_J$ with the value obtained from the exact angular-momentum projection (see Table~\ref{tab:sigma}).
We find very good agreement between the two approaches, confirming that the uncertainty principle provides a consistent interpretation of the projected spin width.
The remaining difference, below 5\%, is consistent with the population of non-$K=0$ components induced by dynamical pair breaking~\cite{Scamps:2023a}.

\begin{figure}[htbp]
    \centering
    \includegraphics[width=\linewidth]{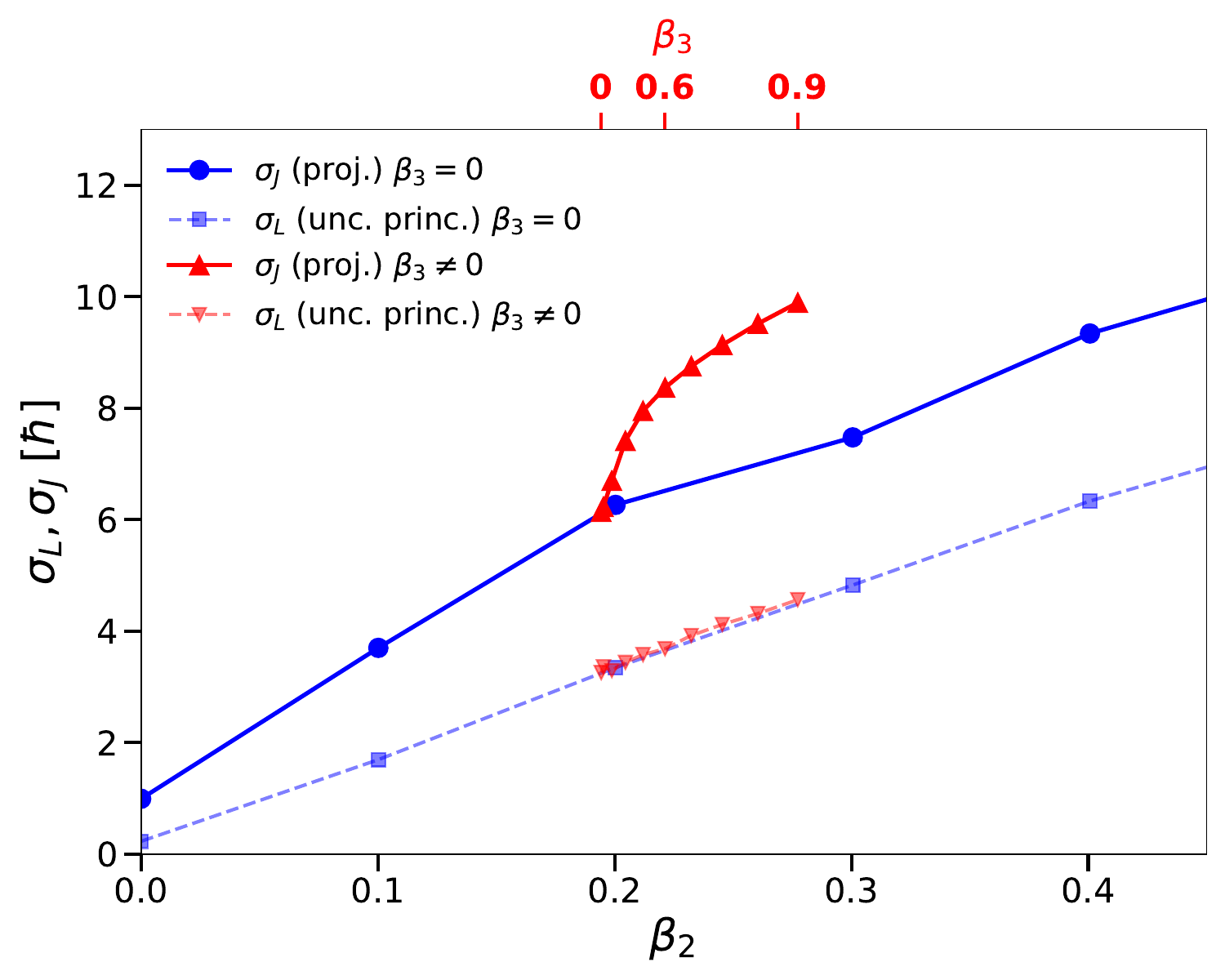}
    \caption{Spin-cutoff parameters $\sigma_J$ (extracted from angular momentum projection) and $\sigma_L$ (calculated via principal axis uncertainty) for $^{144}$Ba as a function of quadrupole deformation $\beta_2$, shown by solid and dashed lines, respectively. Blue curves represent the symmetric case ($\beta_3=0$), while red curves show the evolution with octupole constraint. The specific values for the octupole parameter $\beta_3$ are indicated on the secondary top x-axis.}
    \label{fig:sigma_beta2_comparison}
\end{figure}

\textit{Discussion}—
This works unfolds the relation between the angular momentum of the fragments and there orientation characteristic in a microscopic framework.
The projection method provides an exact results without straightforward interpretation.
Assuming that the overlap \eqref{Gausian_overlap} between rotated many-body states are close to Gaussian gives
accurate results up to a few percents. It also link the angular momentum fluctuation to the width $\sigma_{\beta}$ of the 
Gaussian overlaps in the form of an uncertainty relation. 
Using our Monte Carlo sampler, we further connect 
this width to the fluctuation of the principal axis of the many-body states.
In the case where the fragments deformation is primarily coming from a large quadrupole deformation,
an uncertainty principle directly linking the principal axis orientation and the angular momentum is a good approximation. In other words, the angular momentum of the nucleus can mostly be deduced from a quantum and macroscopic picture where the only degree of freedom of the nucleus is its orientation.
This picture does not hold when the fragments are less deformed and possess a non-vanishing octupole deformation.
In such situation, the distribution of probability for the principal axis orientation is very flat and
the macroscopic picture leads to a negligible angular momentum. In contrast other structural effects (\textit{e.g.} the octupole deformation)
can lead to a small overlap width $\sigma_{\beta}$ and therefore a large spin fluctuation. This effect is illustrated in Fig.~\ref{fig:sigma_beta2_comparison}, where the spin cut-off parameter is shown as a function of quadrupole deformation for an isolated $^{144}$Ba fragment in its lowest-energy deformed state, computed with the same functional and numerical setup as in the previous sections.
In such cases, a purely macroscopic picture linking orientation and spin still gives a correct order of magnitude but significantly underestimate the spin-cutoff. 
When the octupole deformation is increased while the quadrupole constraint is relaxed, the exact projection yields a larger spin cut-off parameter, 
whereas the Monte Carlo estimate follows the same dependency with the $\beta_2$ parameter. 
This demonstrates that the geometric sampling approach misses the contribution of higher-order deformations to the intrinsic angular momentum.

\textit{Concluding remarks}— We have interpreted in details the link between the deformation and the spin of the fission's fragments. 
Greater fragment deformation leads to a stronger alignment with the fission axis.
In the cases of large elongations, an uncertainty relation between the orientation of the principal axis and the angular momentum holds.
A more robust, though less intuitive, uncertainty principle relates the angular momentum fluctuation to the width of the of the overlaps between rotated many-body states. 
As shown in Table~\ref{tab:sigma}, the spin cut-off parameter is reproduced with better than 5\% accuracy by assuming a Gaussian overlap and applying this uncertainty principle.

Thus, the microscopic origin of fragment spin is consistent with the macroscopic–quantal picture~\cite{MIKHAILOV:1999,Bonneau2007,Scamps:2022,Scamps:2023,Shneidman:2025,Shneidman:2002}. Through a complex interplay between the structure of the fissioning nucleus, the properties of the emerging fragments, the nucleus–nucleus interaction, and the Coulomb repulsion, the fragments emerge deformed and oriented at scission. This orientation, through the uncertainty principle, gives rise to angular momentum.
Additional mechanism like the breaking of pair \cite{Scamps:2023a} in the fragments or the coulomb excitation \cite{Wilhelmy:1972,Scamps:2022,Scamps:2023,Randrup:2023} should also contribute to the total spin of the fragments.
 
Finally, although the present approach is fully dynamical, the qualitative features of the results would remain similar in static, constraint-based calculations \cite{Marevic:2021,Bertsch:2019,marevic:2025}. In the TDDFT framework, the dynamical evolution primarily serves to excite the fragments and break nucleon pairs, thereby generating non-zero $K$ components of the spin. The inclusion of dynamics enriches the structure of the fragment angular momentum but does not alter its fundamental origin, which remains governed by the uncertainty principle.

The present results provide a microscopic foundation for macroscopic descriptions of angular-momentum generation in fission, where fragment spin is linked to orientation fluctuations. A more refined description of angular-momentum generation should further investigate the microscopic origin and dynamics of these fluctuations using beyond-mean-field approaches such as time-dependent RPA or time-dependent GCM \cite{simenel2025nuclear}, which naturally extend the present framework.


%

%


\begin{acknowledgments}
We gratefully acknowledge support from the CNRS/IN2P3 supercomputer Center (Lyon, France) for providing computing and data-processing resources needed for this work. This work was also granted access to the HPC resources of IDRIS and CINES under Allocation No. 2024-AD010515531R1 made by GENCI.
\end{acknowledgments}









\bibliography{local_fission.bib}


\end{document}


\preprint{APS/123-QED}



\author{G. Scamps}
\author{A. Guilleux}
\affiliation{Université de Toulouse, CNRS/IN2P3, L2IT, Toulouse, France}

\author{D. Regnier}
\author{A. Bernard}
\affiliation{CEA, DAM, DIF, 91297 Arpajon, France}
\affiliation{Université Paris-Saclay, CEA, LMCE, 91680 Bruyères-le-Châtel, France}


\date{\today}

\title{Supplemental Material : Uncertainty Principle and Angular Momentum Generation in Microscopic Fission Models}


\begin{abstract}
Here we provide some pertinent details to the main text.
\end{abstract}

\maketitle


\section{Constrained HFB initialization and TDHFB time evolution}

The initial configurations used for the dynamical calculations are obtained
from constrained Hartree--Fock--Bogoliubov (CHFB) solutions
computed with the Gogny D1S interaction. The CHFB problem is solved
iteratively by direct diagonalisation in the hybrid single-particle
basis. This basis consists of two-dimensional harmonic oscillator
states restricted to $n_x + n_y \leq N_{\text{shell}}$ with
$N_{\text{shell}} = 9$ and a one-dimensional mesh along the $z$ axis.
The number of quasiparticle states is entirely fixed by the basis size, with
$N_{\text{qp}} = N_{\text{basis}} = 5720$.

\begin{figure}[t]
\centering
\includegraphics[width=0.99\linewidth]{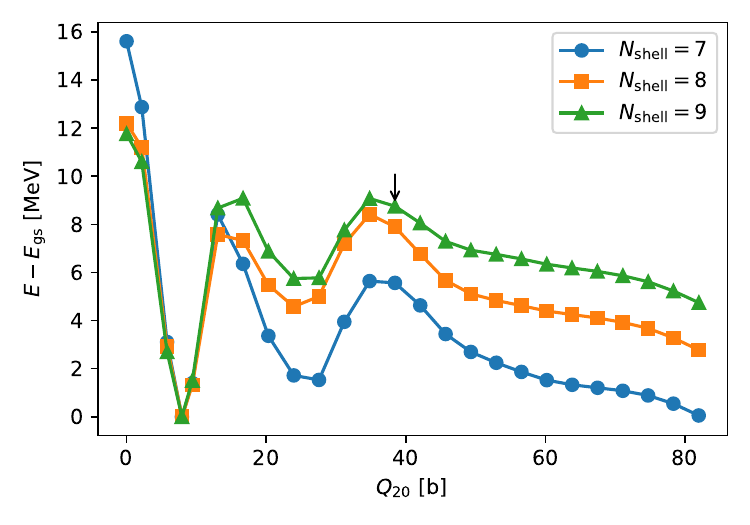}
\caption{Potential energy surface for $^{230}$Th computed with CHFB
with quadrupole and octupole constraints for different $N_{\text{shell}}$ values. The arrow show the starting HFB state of the TDHFB calculation. }
\label{Th230_fission_potential}
\end{figure}

To locate the asymmetric fission valley, HFB states are first generated
with simultaneous quadrupole and octupole constraints. Once the system
reaches the asymmetric valley, the octupole constraint is released and only
the quadrupole moment is kept. This produces a sequence of configurations
along the elongation coordinate. The fission path  is illustrated in Fig.~\ref{Th230_fission_potential} for different values of $N_{\text{shell}}$.  

No spatial symmetry is imposed during the CHFB or TDHFB stages. The hybrid
basis allows unrestricted shapes along the fission axis. Time-reversal
symmetry is broken in the TDHFB evolution.

All quasiparticle states associated with the hybrid basis are included.
There is no additional quasiparticle cutoff. The evolution is performed in
a finite box with fixed (Dirichlet) boundary conditions. No absorbing layer
is added. The time evolution is stopped before any fragment reaches the
edges of the box to avoid reflections.

Although the convergence of the potential energy surface with respect to
the basis size is not fully achieved for $N_{\text{shell}} = 9$, this value
represents the current computational limit of the hybrid-basis
implementation with the Gogny interaction. Increasing $N_{\text{shell}}$
would require a prohibitive number of quasiparticle states and memory
resources. The present configuration already involves
$N_{\text{qp}} = N_{\text{basis}} = 5720$ states and a total computational
cost of about $10^4$~CPU~hours per fission trajectory, using MPI
parallelisation. Despite this limitation, the deformation energy landscape
obtained is sufficiently accurate to describe the qualitative features of
the fission path and the dynamics up to scission. Increasing the number of
shells would not modify the conclusions of the present work.

The dynamical evolution is continued until the fragments reach a separation
distance of 6\,fm between their skin. Once the fragments are well separated, fragment
properties are extracted. These include the mass, charge, total kinetic
energy, and the saddle-to-scission time. A summary of these quantities is
provided in Table~\ref{tab:fragment_observables} for
$^{230}$Th, $^{240}$Pu, and $^{250}$Cf.

\begin{table}[t]
\centering
\caption{Properties of fragments obtained from the TDHFB calculations.}
\label{tab:fragment_observables}
\begin{tabular}{l||c|c|c|c|c|c}
\hline\hline
Nucleus & $A_L$ & $A_H$ & $Z_L$ & $Z_H$ & TKE [MeV] & $T_{\text{s\textendash s}}$ [zs] \\
\hline
$^{230}$Th & 87.5 & 142.5 & 35.0 & 55.0 & 153 & 12.3 \\
$^{240}$Pu & 105.3 & 134.7 & 41.9 & 52.1 & 186 & 6.5 \\
$^{250}$Cf & 112.6 & 137.4 & 44.4 & 53.6 & 178 & 9.6 \\
\hline\hline
\end{tabular}
\end{table}

\begin{figure}[t]
\centering
\includegraphics[width=0.99\linewidth]{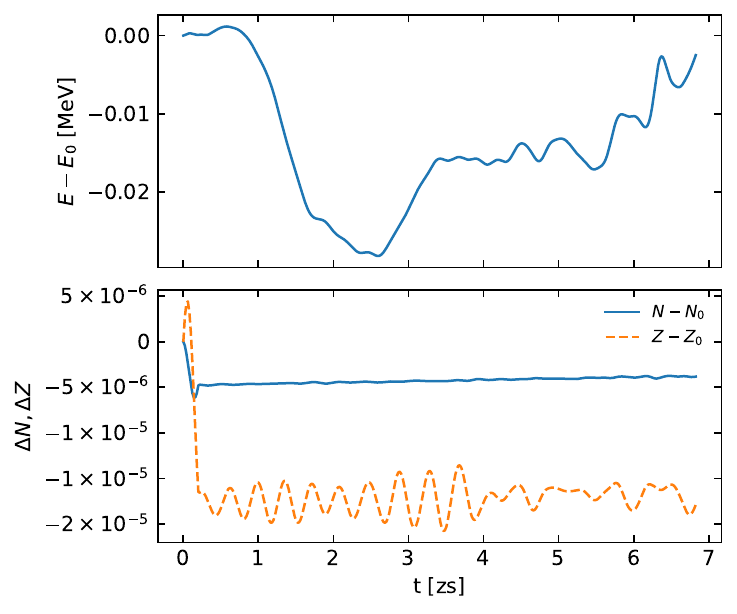}
\caption{Conservation of total energy (top) and particle number (bottom) during the
TDHFB time evolution.}
\label{conservation}
\end{figure}

The conservation of particle number and total energy is checked throughout
the TDHFB time evolution. In Fig.~\ref{conservation}, the numerical error
in both quantities remains very small during the trajectory.

\section{Preparation of the wave function}

Once the fragments are well separated, the TDHFB wave function in the hybrid basis is transformed to a convenient representation for further analysis. First, a Bloch–Messiah decomposition is performed on the reduced density matrices of protons and neutrons. For each isospin $\tau$ (protons or neutrons), the reduced density matrix $\rho_\tau$ is constructed from the Bogoliubov transformation matrices $V^{(\tau)}$ as
\begin{align}
    \rho_\tau(i,j) = \sum_k V_{ik}^{(\tau)} V_{jk}^{(\tau)*},
\end{align}
and subsequently diagonalized as
\begin{align}
    \sum_j \rho_\tau(i,j) \, \varphi_{\alpha}^{(\tau)}(j) 
    = n_\alpha^{(\tau)} \, \varphi_{\alpha}^{(\tau)}(i),
\end{align}
where $n_\alpha^{(\tau)}$ are the canonical quasiparticle occupation numbers and $\varphi_\alpha^{(\tau)}(i)$ the corresponding canonical states, expressed on the hybrid basis labeled by $i = (n_x, n_y, i_z)$.  
The canonical partner of $\varphi_\alpha^{(\tau)}(i)$ is denoted by $\varphi_{\bar{\alpha}}^{(\tau)}(i)$.

The anomalous density in the canonical basis is computed as  
\begin{align}
    \kappa_\alpha^{(\tau)} 
    = \sum_{i,j} 
      \varphi_{\alpha}^{(\tau)*}(i) \, 
      \tilde{\kappa}_{ij}^{(\tau)} \, 
      \varphi_{\bar{\alpha}}^{(\tau)}(j),
\end{align}
with  
\begin{align}
    \tilde{\kappa}_{ij}^{(\tau)} 
    = \sum_{k} V_{ik}^{(\tau)*} \, U_{jk}^{(\tau)}.
\end{align}

Due to numerical errors, the magnitude of $\kappa_\alpha^{(\tau)}$ obtained from this projection is not exactly equal to the expected value $\sqrt{n_\alpha^{(\tau)} \left( 1 - n_\alpha^{(\tau)} \right)}$. A subsequent normalization is applied to enforce the correct magnitude, ensuring consistency with the canonical occupation numbers. We also enforce the properties
\begin{align}
    n_{\bar{\alpha}} &= n_{\alpha}, \\
    \kappa_{\bar{\alpha}} &= -\,\kappa_{\alpha},
\end{align}
which are also slightly broken due to the numerical errors during the time propagation.

After the Bloch–Messiah decomposition and normalization, the canonical states $\varphi_\alpha^{(\tau)}(i)$ are projected onto a three-dimensional Cartesian grid of size $n_x \times n_y \times n_z = 30 \times 30 \times 52$ with a mesh spacing of $\Delta x = 0.8$ fm.

\section{Nucleoscope}

\subsection{Sampling of Many-Body Configurations}

To go beyond the one-body density picture, the orientation of each fragment is
extracted from many-body configurations sampled from the Bogoliubov vacuum.  
These configurations are generated using the \textsc{Nucleoscope} code, which employs
a Markov Chain Monte Carlo algorithm to sample the joint probability distribution of
the nucleons positions and intrinsic spins.
The sampler relies on multiple independent Metropolis Markov chains
characterized by a Gaussian spatial transition kernel distribution with a width
$1.5$~fm and a spin-flip proposal probability of $0.1$.
The metropolis algorithm requires estimations of ratio of probabilities
for the nucleons probabilities which can be evaluated from the canonical
representation of the Bogoliubov vacuaa following Ref.~\cite{matsumoto2022visualization}.
In practice, we generate a total $160\times 10^6$ Monte Carlo events for each fissioning system
out of 32 independent Markov chains.
We remove from this raw sample, the first $10^5$ events of each chain to account for the warm up time of the
Markov chains.
To mitigate the auto-correlation within each Markov chain, we finally retain only one event out of 50.
We end up with a sample of more than $3\times10^6$ events used to actually compute observables.

For each event, the fragments are identified, and only fragment mass numbers
$A_F$ with at least $100$ sampled occurrences are retained for the final
$\theta$-angle analysis as a function of $A$ presented in the main text.

\begin{figure*}[t!]
    \centering
    \includegraphics[width=0.49\textwidth]{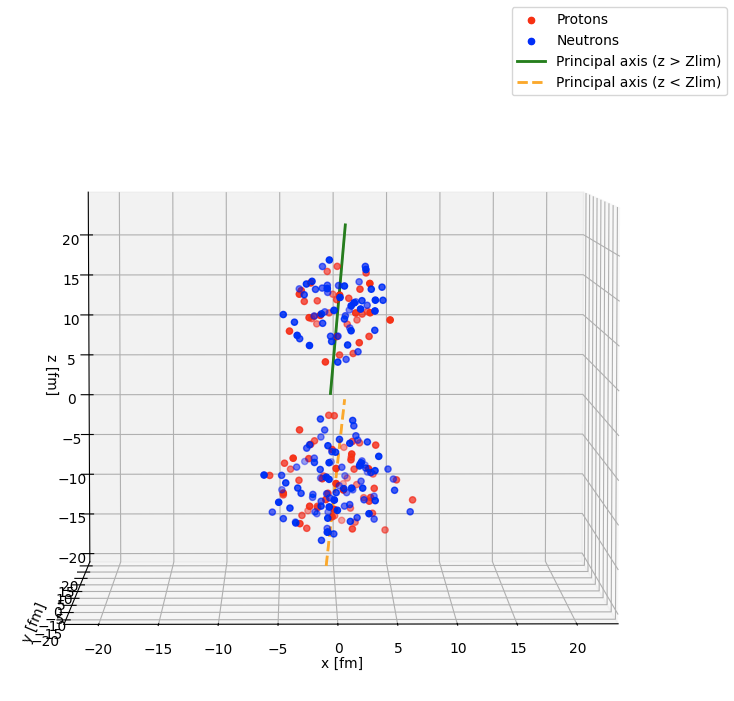}
    \includegraphics[width=0.49\textwidth]{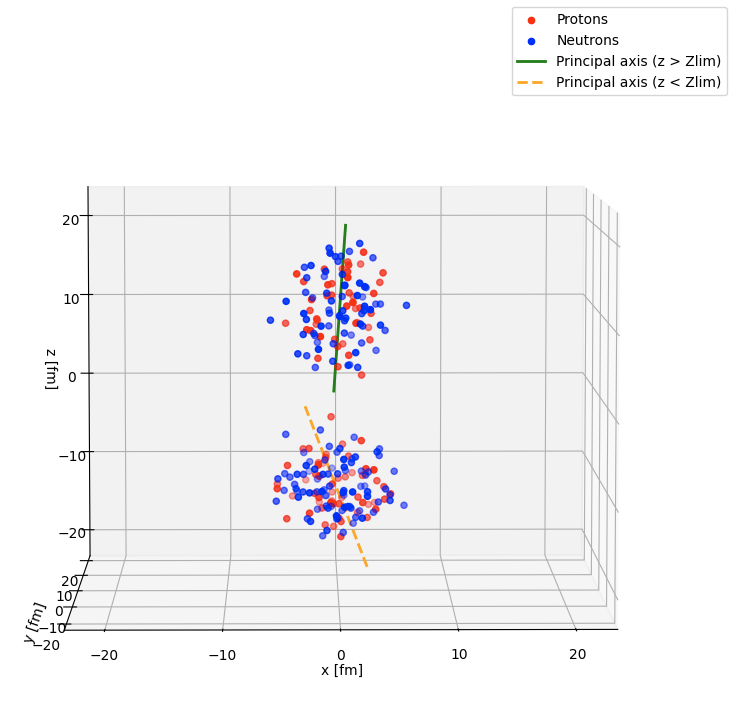}
    \caption{
        Example of the determination of the principal deformation axes
        for two fissioning systems.
        The left panel corresponds to $^{230}\mathrm{Th}$, and the right panel
        to $^{250}\mathrm{Cf}$. 
        The red and blue dots represent the proton and neutron spatial distributions, respectively. 
        The green and orange lines indicate the principal axes
        of the fragments located above ($z > Z_{\text{lim}}$) 
        and below ($z < Z_{\text{lim}}$) the separation plane. 
        The angle $\theta$ is then compute as  the angle between each fragment’s
        principal symmetry axis and the global $z$-axis of the fission coordinate system.
    }
    \label{fig:principal_axis_example}
\end{figure*}

\subsection{Determination of the Angle $\theta$}

For each event, we compute the principal deformation axis of each fragment
using the quadrupole tensor
\begin{align}
Q_{ij} = \sum_{k} \left( 3 r_{k,i} r_{k,j} - \delta_{ij} r_k^2 \right),
\end{align}
where $\vec{r}_k$ denotes the position of nucleon $k$ relative to the fragment's center of mass.
The eigenvector corresponding to the largest eigenvalue of $Q_{ij}$ defines the
principal symmetry axis of the fragment, denoted by the unit vector $\hat{v}$.

The orientation of this axis with respect to the fission axis (chosen as the $z$-axis)
is characterized by the polar angle $\theta$, defined as
\begin{align}
\theta = \arccos(\hat{v} \cdot \hat{z}).
\end{align}

If the octupole moment $Q_{30}$ of the fragment is negative,
the axis direction is inverted ($\hat{v} \rightarrow -\hat{v}$)
to ensure a consistent orientation. Two examples of determination of the principal axis are shows on fig.~\ref{fig:principal_axis_example}
The resulting distribution of $\theta$ reflects the angular alignment of
the fragment’s intrinsic deformation axis relative to the fission axis.

\section{Angular momentum projection}

After the wave functions are prepared on a three-dimensional Cartesian grid, the projection onto angular momentum states is performed to extract the spin distributions of the fission fragments. 
The projection operator onto angular momentum $J_F$, $K_F$ is performed by discretizing the triple integral:

\begin{widetext}
\begin{equation}
P(J_F,K_F) = \frac{2J_F+1}{16\pi^2} \int_0^{2\pi} d\alpha \int_0^{\pi} d\beta \, \sin\beta \int_0^{4\pi} d\gamma \, {\cal D}^{J_F*}_{K_F K_F} (\alpha, \beta, \gamma) \,
\langle \Psi | e^{i\alpha \hat J_z} e^{i\beta \hat J_y} e^{i\gamma \hat J_z} | \Psi \rangle,
\label{eq:PJMK}
\end{equation}
\end{widetext}

the $\alpha$ and $\gamma$ integral are discretized with a uniform spacing of $\pi/8$.

The angle $\beta$ is discretized using a  mapping to account for the $\sin\beta$ integration weight. A uniform grid in the variable $\mu = \arccos(1 - 2\beta/\pi)$ is used, corresponding to a Gauss-Legendre quadrature in $\beta$ (as in Ref.~\cite{DOBACZEWSKI20092361}), which produces denser sampling near $\beta = 0$ and $\beta = \pi$. In the present calculations, the integral over $\beta$ is approximated with $N_\beta = 32$ points.


The rotation of the wave function is performed directly in coordinate space using a cardinal sine (\textit{sinc}) interpolation method. For each Euler angle, the spatial components of the quasiparticle wave functions are rotated around a selected axis by an angle $\theta$. The rotation is implemented by evaluating the transformed coordinates
\begin{align}
(x',y',z') &= 
R(\theta)\,(x - x_{\text{cm}},\, y - y_{\text{cm}},\, z - z_{\text{cm}}) 
\nonumber \\[4pt]
&\quad + (x_{\text{cm}},\, y_{\text{cm}},\, z_{\text{cm}}),
\end{align}
and reconstructing the rotated wave function on the numerical grid using cardinal sine functions:
\begin{widetext}
\begin{align}
f(x',y',z') \approx 
\sum_{ijk} f(x_i,y_j,z_k)
\, \mathrm{sinc}\!\left(\pi\frac{x'-x_i}{\Delta x}\right)
\, \mathrm{sinc}\!\left(\pi\frac{y'-y_j}{\Delta x}\right)
\, \mathrm{sinc}\!\left(\pi\frac{z'-z_k}{\Delta x}\right),
\end{align} 
\end{widetext}

where $\mathrm{sinc}(x) = \sin(x)/x$.  
This approach preserves the spatial accuracy and continuity of the wave function inside a cutoff radius $r_{\text{cut}}$=10~fm from the fragment center of mass. Before performing the projection, the center-of-mass momentum of each fragment is removed by applying a compensating Galilean boost to the corresponding wave function, ensuring that the rotation is carried out in the fragment rest frame.

In addition to the spatial rotation, the spin components of each quasiparticle
state are consistently rotated using the SU(2) spinor transformation.
Each quasiparticle wave function carries a two–component spinor
\begin{align}
\begin{pmatrix}
\varphi_{\uparrow} \\
\varphi_{\downarrow}
\end{pmatrix},
\end{align}
which transforms under a rotation by an angle $\theta$ around the axis $\hat{n}$ as
\begin{align}
\begin{pmatrix}
\varphi_{\uparrow}' \\
\varphi_{\downarrow}'
\end{pmatrix}
=
\exp\!\left( -\,\frac{i\,\theta}{2}\, \boldsymbol{\sigma}\!\cdot\!\hat{n} \right)
\begin{pmatrix}
\varphi_{\uparrow} \\
\varphi_{\downarrow}
\end{pmatrix},
\end{align}
where $\boldsymbol{\sigma}$ denotes the Pauli matrices.
The combined spatial and spin rotations yield the rotated quasiparticle wave functions
\begin{align}
|\Psi(\theta)\rangle = \hat{R}(\theta)\,|\Psi\rangle,
\end{align}
which are then used in the angular momentum projection integral in
Eq.~(\ref{eq:PJMK}).

The overlap between the initial state and the rotated one is compute with the pfaffian method using the algorithm of Ref.~\cite{Wim12}

\section{Gaussian overlap and fluctuation of the angle}

Although the $\theta$ angle operator is not explicitly defined, a state localized in the collective orientation angle $\theta$ can be formally represented as a superposition of orientation eigenstates:
\[
|\psi\rangle
=
\int d\theta \,
f(\theta)\,|\theta\rangle.
\]

Rotation operators act as translations in the collective coordinate $\theta$, i.e.
\[
R(\alpha)\,|\theta\rangle = |\theta+\alpha\rangle,
\]
which reflects the definition of $\theta$ as an orientation angle in the collective space.

If we consider a Gaussian wave packet,
\[
f(\theta)
=
\frac{1}{(2\pi\sigma^2)^{1/4}}
\exp\left(
-\frac{\theta^2}{4\sigma^2}
\right),
\]
which characterizes fluctuations of the orientation angle with width $\sigma$.

The overlap under a rotation $R(\alpha)=e^{-i\alpha \hat J}$ is then
\[
\langle\psi|R(\alpha)|\psi\rangle
=
\exp\left(
-\frac{\alpha^2}{8\sigma^2}
\right).
\]

This result shows that Gaussian fluctuations in the collective orientation coordinate lead directly to a Gaussian decay of the rotational overlap kernel. The width of this kernel is therefore determined by the fluctuation amplitude $\sigma$, reflecting the conjugate nature of orientation and angular momentum and setting the scale of angular-momentum fluctuations.

%









\bibliography{local_fission_SM.bib}
